\documentclass{bmvc2k}

\usepackage{amsmath,amssymb, amsfonts,graphicx,color,comment, soul, multirow, rotating, enumerate, multirow}

\DeclareMathOperator*{\argmin}{\textbf{arg\,min}}
\newcommand\blfootnote[1]{%
  \begingroup
  \renewcommand\thefootnote{}\footnote{#1}%
  \addtocounter{footnote}{-1}%
  \endgroup
}

\newcommand{\PG}[1]{\textcolor{black}{#1}}
\newcommand{\Isa}[1]{\textcolor{black}{#1}}
\newcommand{\HBO}[1]{\textcolor{black}{#1}}

\newcommand{\GLB}[1]{\textcolor{black}{#1}}

\title{Anatomically constrained CT image translation for heterogeneous blood vessel segmentation}

\addauthor{Giammarco La Barbera}{giammarco.labarbera@telecom-paris.fr}{1}
\addauthor{Haithem Boussaid$^{\ast}$}{}{6,2}
\addauthor{Francesco Maso}{}{1}
\addauthor{Sabine Sarnacki}{}{3,4}
\addauthor{Laurence Rouet}{}{2}
\addauthor{Pietro Gori}{}{1}
\addauthor{Isabelle Bloch}{}{5,1,3}
\addinstitution{
LTCI\\ T\'el\'ecom Paris, Institut Polytechnique de Paris\\ France
}
\addinstitution{
Philips Research Paris\\ Suresnes, France
}
\addinstitution{
IMAG2\\ Institut Imagine, Universit\'e Paris Cité\\ France
}
\addinstitution{
Universit\'e Paris Cité\\ Hôpital Necker Enfants-Malades, APHP\\ France
}
\addinstitution{
Sorbonne Universit\'e\\ CNRS, LIP6\\ Paris, France
}
\addinstitution{
Technology Innovation Institute\\ Abu Dhabi, UAE
}

\runninghead{La Barbera \etal}{Anatomically constrained CycleGAN}

\def\etal{\emph{et al}\bmvaOneDot}

\begin{document}

\maketitle
\blfootnote{$^{\ast}$This work was done while the co-author was with Philips Research Paris and is now affiliated with Technology Innovation Institute, Abu Dhabi, UAE}
\begin{abstract}
Anatomical structures \HBO{such as blood vessels} in 
contrast-enhanced CT (ceCT) images can be \HBO{challenging} to segment due to 
the variability in contrast medium diffusion. 
The combined use of ceCT and contrast-free (CT) CT images 
{can improve the segmentation performances}, but at the cost of a double radiation exposure. 
To limit the radiation dose, generative models could be used to synthesize one modality, instead of acquiring it.
\HBO{The CycleGAN approach has recently attracted particular attention because it alleviates the need for paired data that are difficult to obtain. 
Despite the great performances demonstrated in the literature, limitations still remain when dealing with 3D volumes generated slice by slice from unpaired datasets with different fields of view.}
We present an extension of CycleGAN to generate \HBO{ high fidelity images, with good structural consistency, \GLB{in this context.}
We leverage anatomical constraints and automatic region of interest selection by adapting the {\em Self-Supervised Body Regressor}. These constraints enforce anatomical consistency and allow feeding anatomically-paired input images to the algorithm. 
Results show qualitative and quantitative improvements, compared to state-of-the-art methods, on the translation task between ceCT and CT images (and vice versa).
{\GLB{Moreover,} using the CT images produced by our algorithm, we achieve blood vessel segmentation performance on par with the segmentation performance using real CT images.}}
\end{abstract}

\section{Introduction}
\label{sec:intro}
Heterogeneity in contrast is one of the major difficulties in medical image segmentation when using Convolutional Neural Networks (CNN), 
in particular in 
constrast-enhanced \HBO{Computed Tomography} (ceCT)
images. 
The effect of the contrast agent on the pixel intensity is not always the same among patients due to different factors, such as acquisition times and patient morphology. 
Furthermore, the presence of a tumor or thrombosis in the vessels can also cause heterogeneity in contrast within an anatomical structure.
This raises difficulties during segmentation, and manual corrections are often needed.

In~\cite{sandfort2019,song2020,zhu2020} the authors show that the combined use of ceCT and contrast-free (CT) CT images is able to deal with 
the heterogeneity of ceCT images and thus improves segmentation. However, in order to limit ionising radiations, clinicians often acquire only one CT modality. 
One common \HBO{computational} approach to compensate for the absence of an imaging modality is to use generative models~\cite{dda-gan,reviewGAN} to synthesise it. In the absence of paired data sets, unsupervised 
translation methods, based on CycleGAN~\cite{zhu2017cyclegan} and UNIT~\cite{unit}, {have been proposed}~\cite{newreviewGAN,reviewGAN,dar2019image,pan2018synthesizing}. Some authors have also already considered applying CycleGAN~\cite{sandfort2019,song2020} or UNIT~\cite{zhu2020} to artificially remove or add contrast medium on CT images.
CycleGAN~\cite{zhu2017cyclegan} is an evolution of Generative Adversarial Network (GAN)~\cite{goodfellow}, which introduces a second neural network that tries to solve the inverse task, namely reconstructing the input. A cycle consistency loss function is combined to the adversarial loss to overcome the lack of paired data. UNIT~\cite{unit} is another model conceived for the unpaired setting. This generative model is composed of two variational autoencoder networks, which work on two different domains but share the same latent space. 
Different modifications have already been proposed for both methods, such as the use of Wasserstein distance~\cite{wassersteinGAN}, attention mechanisms~\cite{Attention,ugatit} and U-Net as discriminator network~\cite{schonfeld2020u}. 
However, 
these models do not \HBO{guarantee} to preserve fine structures~\cite{zhu2020} and may produce artefacts~\cite{minkyo2021,reviewGAN}, which prevent {their use for the segmentation of small and heteregoneous structures, such as blood vessels.}
\GLB{In particular, the cycle consistency loss function enforces a relationship only at a distribution level.} \HBO{In \cite{Moriakov2020}, the authors demonstrate that CycleGAN can deliver ambiguous solutions,} especially for substantially different distributions as in medical imaging. Several works tried to address this limitation by adding more terms to the loss function, such as mutual information~\cite{MI2019} and perceptual loss term~\cite{perceptual,ploss}, that require no supervision. Despite the different methods proposed, the anatomical constraint remains insufficient.


As a matter of fact, \HBO{another challenge when dealing with unpaired 3D medical images 
is the lack of 3D consistency. With current hardware memory limitations, it is difficult to train a 3D network taking as input a whole 3D volume. Instead, a common approach is to use 2D networks that take a slice of a 3D volume along \GLB{one} axis.} 
Moreover, in the unpaired scenario we can have different number of slices for the same anatomical region among patients, leading to difficulties to select anatomically-paired slices.
In fact, in~\cite{minkyo2021}, authors showed that it is fundamental to inform the generator on the specific regions that should be affected by the contrast materials. For this reason, the use of \textit{not-aligned} paired data is more effective than unpaired data. Some authors~\cite{pbs} claim that the use of unpaired data can be mitigated by exploiting the approximately common anatomy between subjects. They refer to this as \textit{position-based selection (PBS)} strategy. 
However, in the abdominal region, the different sizes and lengths of the organs must be taken into account, implying that the slice $n$ of the patient $i$ with $N$ slices may not have the same anatomical content as slice $m = n \cdot \frac{M}{N}$ of patient $j$ with $M$ slices.
Eventually, the use of 3D {\em affine} registration (\textit{e.g.,} as in the Simple-Elastix~\cite{simple-eslatix} library) could be a solution to the problem, but the difference between the two domains, the difficulty of identifying the fixed reference image, and the high variability in shape and relative size and pose of abdominal organs among subjects (especially in 3D) may lead to misalignment. 

To address these issues, we propose an extension of the CycleGAN method which includes:
\begin{enumerate}[(i)]
 \setlength{\itemsep}{1pt}
 \setlength{\parskip}{0pt}
 \setlength{\parsep}{0pt}
 \item the automatic selection of the region of interest by exploiting anatomical information, in order to reduce the anatomical distribution of 3D data \HBO{acquired with different fields of view;}
 \item the use of a {\em Self-Supervised Body Regressor} (SSBR), \HBO{adapted from}~\cite{ssbr}, to select anatomically-paired slices among the unpaired ceCT and CT domains, {and} help the discriminator to specialize in the task; 
 \item the use of the SSBR score as an extra loss function that constrains the generator to produce a slice describing the same anatomical content as the input, \HBO{inspired from} the auxiliary classifier GAN~\cite{acgan};
 \item the use of the input image as a template for the generator, as in~\cite{paired2018}, and the use of an anatomical binary mask to constrain the output.
\end{enumerate}

The proposed method is generic and could be used in \GLB{different medical applications, i.e. different body regions such as brain or lungs, or different translation modalities such as MRI to CT or T1-w to T2-w}. Here, we propose to use it for the generation of CT abdominal images from ceCT images and vice versa. We test the use of a generated modality, in combination with the complementary original one, 
to improve segmentation performance on blood vessels of pathological patients. 
{To the best of our knowledge, this strategy has never been tested for such an application.}

We show that our method greatly improves the ceCT-CT translation quality compared to state-of-the-art methods. As a consequence, the segmentation performances using generated images are also improved, achieving both qualitative and quantitative results comparable to 
{the ones using} both real images.

It is important to highlight that, in this work, the use of synthetic images is intended 
{to increase 
segmentation performances} and {not} to use for clinical diagnosis. 

\GLB{It should be also noted that in this paper we focus on CNN-based methods, as state-of-the-art methods for unsupervised medical image translation. While interesting works on image-to-image translation based on $transformers$ are starting to be explored~\cite{transgan,instaformer}, their application in the medical domain is limited due to the restricted number of data available. For this reason, existing works focus only on paired medical data sets~\cite{resvit}. Nevertheless, for the sake of completeness, we test a transformer-GAN, namely TransGAN~\cite{transgan}, on our unsupervised medical task.}

\section{Proposed Method}
\label{sec:meth}

\begin{figure}[htbp]
\begin{center}
\includegraphics[width=\linewidth]{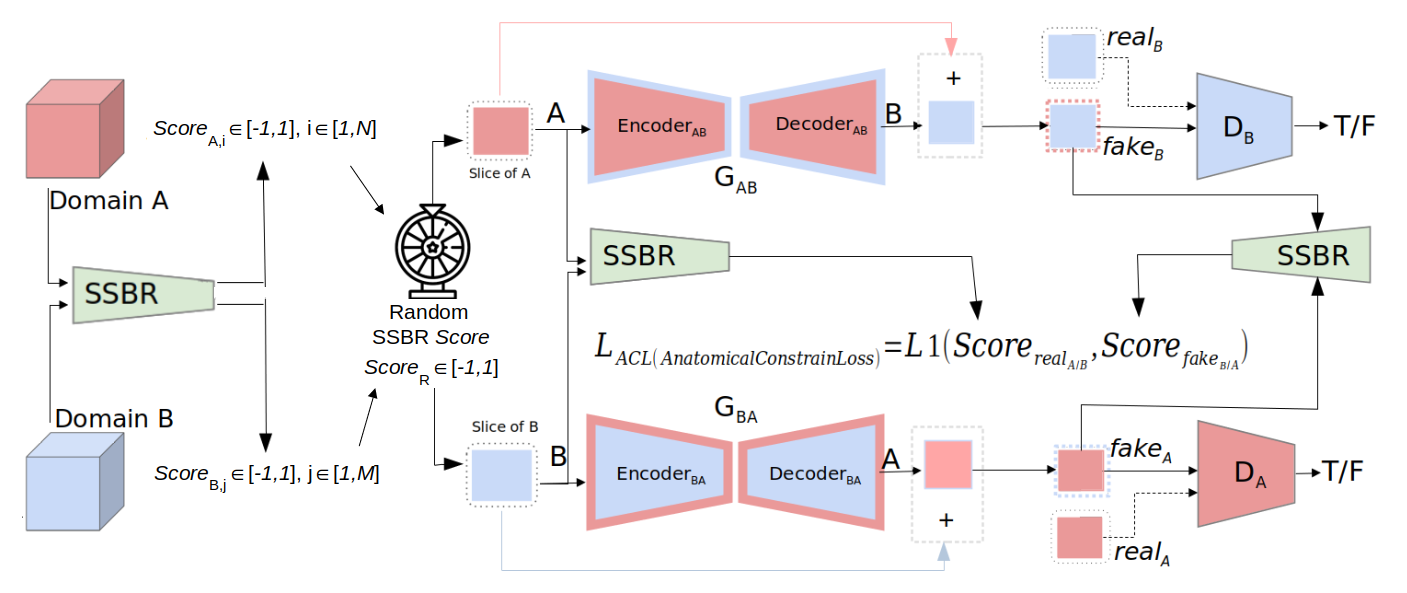}
\end{center}
\caption{Proposed method for the selection of anatomically-paired slices via {\em Self-Supervised Body Regressor}, and its use as a loss function $L_{ACL}$.}
\label{fig:pipeline}
\end{figure}

\noindent As stated in Section~\ref{sec:intro}, when using 2D unpaired medical data, the selection of consistent (\textit{i.e.,} corresponding to the same region of interest (ROI)) and anatomically similar slices between the two domains is highly important to facilitate the generative process.
To this end, we propose 
to leverage a Self-Supervised Body Regressor (SSBR)~\cite{ssbr}, 
a CNN that finds common features on anatomically similar slices from unlabeled CT images. 
This results in assigning the same label for slices describing the same anatomy while belonging to different patients. The SSBR is trained to estimate slice scores
which are monotonous functions of 
the slice indices.
However, there is no guarantee to obtain the same range of scores for different modalities. We propose a solution to this problem.
The method described in this section is summarized in Figure~\ref{fig:pipeline}.

\textbf{Input selection via SSBR}
\GLB{First of all, because of the different fields of view (FOV) in the two datasets, it is important to 
\Isa{select an appropriate} ROI.}
This can be done off-line and manually, as in~\cite{sandfort2019,zhu2020}. Here, instead, we first propose a simple, automatic and on-line method {to select only slices from the abdominal region}.
We automatically select the first slice of the lungs and the last slice of the intestinal area as upper and lower landmarks, which is easy due to the strong presence of black pixels in both ceCT and CT acquisitions.

Then, instead than PBS strategy as in~\cite{pbs}, we propose the use of an SSBR, as shown in Figure~\ref{fig:pipeline}.
For training, we optimize three loss functions \GLB{that do not require annotated anatomical labels}. 
The first one, as in~\cite{ssbr}, favors an increasing order of SSBR scores according to the positions of the slices, \GLB{avoiding repeating scores and ensuring similar scores for adjacent regions}:
\begin{equation}
 L_{order} = - \sum_{k=1}^K \sum_{p=1}^{P-1} \log (h (Score_{k,p+1} - Score_{k,p}))
\label{eq:order}
\end{equation}
where $Score_{k,p} \Isa{\in [-1,1]}$ is the SSBR output for slice $p$ of CT volume $k$, $h$ is the sigmoid activation function, $K$ is the number of CT volumes in the chosen set (mini-batch) and $P$ is the number of slices in each volume. 

The second loss function exploits the automatic selection of the ROI, forcing the first and last slices to have a score of -1 and +1 respectively:
\begin{equation}
 L_{norm} = \sum_{k=1}^K (f(Score_{k,1} + 1) + f(Score_{k,P} - 1))
\label{eq:norm}
\end{equation}
where $f$ is a smoothed L1 norm. {This function guarantees the same score range for both modalities.}

The third loss function takes into account the anatomical variability of the abdominal area. Using the binary mask $BM$ of the body for each slice (easily obtained in CT), we want the difference between successive scores to be an increasing function of
the normalized cardinality 
of the intersection of the $BM$ of successive slices: 
\begin{equation}
\begin{gathered}
 L_{anat} = \sum_{k=1}^K \sum_{p=1}^{N-1} f (\Delta^{BM}_{k,p+1} - \Delta_{k,p+1})) 
 \\
 \mbox{with } \Delta^{BM}_{k,p}= 1 - \frac{|BM_{k,p} \cap BM_{k,p-1}|}{|BM_{k,p-1}|}
 \mbox{ and } \Delta_{k,p}=Score_{k,p}-Score_{k,p-1}
\label{eq:anat}
\end{gathered}
\end{equation}
\GLB{This is done in order to increase the difference in score between slices with higher anatomically difference and not fall into the trivial linear solution.}

Eventually, the terms of the cost function are combined by a weighted average, and the function to be optimized is:
\begin{equation}
 L_{SSBR} = \alpha L_{order} + \beta L_{anat} +\gamma L_{norm}
\label{eq:ssbr}
\end{equation}
where $\alpha$, $\beta$ and $\gamma$ are empirically chosen weights that balance the three losses.

Once the SSBR is properly trained \GLB{(details in the next section),} to extract the anatomically-paired slices for each iteration of the CycleGAN we do the following:
\begin{enumerate}
 \setlength{\itemsep}{2pt}
 \setlength{\parskip}{0pt}
 \setlength{\parsep}{0pt}
 \item A single patient is selected for each of the unpaired ceCT and CT domains, called domains A and B;
 \item The 3D volumes are automatically restricted to the abdominal region;
 \item SSBR scores are predicted for each 2D slice of the two 3D ROIs, using the pre-trained SSBR;
 \item $J$ random SSBR scores, denoted by $Score_{R_j}$,  are sampled in $[-1,1]$, where~$J$ is the selected number of slices corresponding to the size of the mini-batch;
 \item For each $Score_{R_j}$, the slice with the closest score is selected in each domain, as $\argmin_{p} |Score_{R_j} - Score_{\cdot,p}|$ where $\cdot$ is the domain (A or B) and $p$ is the selected slice in $[1,N]$ for A and $[1,M]$ for B.
\end{enumerate}

\textbf{Anatomically constrained CycleGAN}
Inspired by~\cite{acgan}, we propose the use of the pre-trained SSBR as an auxiliary classifier to enforce the \PG{anatomical} consistency (i.e., same body parts) between the input and the synthesized output. During the training phase of the generator, we add to the loss functions of the standard CycleGAN an L1 norm between the SSBR score of the input $real$ A (resp. $real$ B) and the SSBR score of the generated slice $fake$ B (resp. $fake$ A), called the {\em Anatomical Constraint Loss} ($ACL$), as shown in Figure~\ref{fig:pipeline}:
\begin{equation}
 L_{ACL}=\frac{1}{J} \displaystyle \sum_{j=1}^J|Score_{real_{A/B},j}-Score_{fake_{B/A},j}|
\label{eq:acl}
\end{equation}

We also propose to further constrain the models in two ways. 
First, as in~\cite{paired2018}, we use the input image as a template ($I_nA_d$ in images and tables), i.e. the generators only need to estimate how to modify the input image without estimating an output image from scratch. 
Secondly, during inference, we remove the artefacts created in the original black areas (e.g. background or air). Here, we use the binary mask $BM$ used also in Equation~\ref{eq:anat}. 

\section{Results and Discussion}
\label{sec:rad}


\subsection{\GLB{Implementation details}} 
The hyperparameters for CycleGAN and UNIT were found empirically on the training set, starting from those in~\cite{zhu2017cyclegan} and~\cite{unit} respectively. The best combination of weights for CycleGAN losses was found as $0.5$ for identity loss, $10$ for cycle-consistency loss and $1$ for both adversarial loss and our $L_{ACL}$ loss function. For KL loss of UNIT we set a weight of $0.01$. 

For the SSBR, we operated as in~\cite{ssbr}, using ResNet-34~\cite{res-net} as the backbone. The best weights in Eq.~\ref{eq:ssbr} were found empirically as: $\alpha=5\cdot10^{-3},\beta=1,\gamma=10$. 

All trainings and tests were \Isa{run} on a GPU NVIDIA\textsuperscript{\tiny {\textregistered}} Tesla\textsuperscript{\tiny {\textregistered}} P100 with 16 GB of VRAM using a mini-batch size of 8.


\begin{figure}[h!]
 \centering
\includegraphics[width=0.7\linewidth]{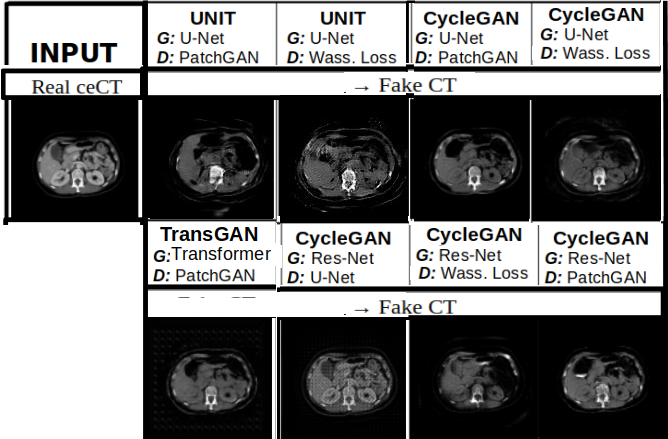}
\includegraphics[width=0.7\linewidth]{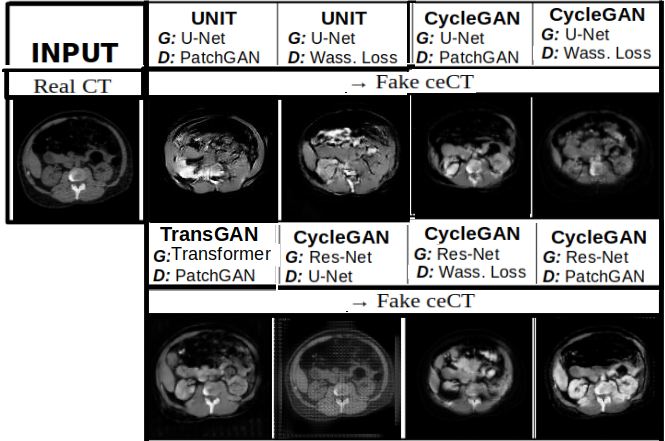}
\caption{Comparison of some state-of-the-art methods on slices of the unpaired test set. Top table: ceCT to CT. Bottom table: CT to ceCT. The slices in all tests are selected with PBS. The input in the other direction gives an idea of what the expected result should look like.}
\label{fig:sota}
\vspace{-4mm}
\end{figure}

\subsection{Qualitative results on unpaired datasets}
\textbf{Dataset} 
For the unpaired image-to-image translation training, the images were obtained from the Cancer Imaging Archive (TCIA)~\cite{tcia}. Two databases of pathology-free abdominal CT images were used: {\em CT Pancreas}~\cite{pancreas-ct} with 82 ceCT images of abdomens, and {\em CT Colonography}~\cite{colon-ct} which contains non-contrast CT images of which 82 healthy subjects were retained for consistency with the first dataset. For both datasets, 72 patients are used for training and 10 for testing. All axial slices were resized from 512$\times$512 to 128$\times$128 pixels to make the training less computationally and memory intensive.\\
\textbf{Experiments}
First, we tested several existing methods~\cite{wassersteinGAN,Attention,unit,zhu2017cyclegan} \PG{to select the best networks for the generators (G) and the discriminators (D)}. The use of the only renal region slices was deemed essential for our experiments to obtain anatomically consistent images and our automatic detection of the abdominal region proved effective, removing the need for manual selection. Some qualitative examples about automatic ROI selection is showed in supplementary materials. Moreover, in all these tests, we used the PBS~\cite{pbs} strategy for selecting slices at the same relative position.
\GLB{Although different methods gave satisfactory results for the easier task of ceCT2CT, only the CycleGAN~\cite{zhu2017cyclegan} with ResNet as the generating network and PatchGAN as the discriminating mechanism produced good results in terms of contrast realness for the task of CT2ceCT, as shown in Figure~\ref{fig:sota}. 
Probably for methods such as TransGAN, performances are limited by the restricted amount of data and computational power.}

\begin{figure}[h!]
 \centering
\includegraphics[width=0.65\linewidth,height=0.43\textheight]{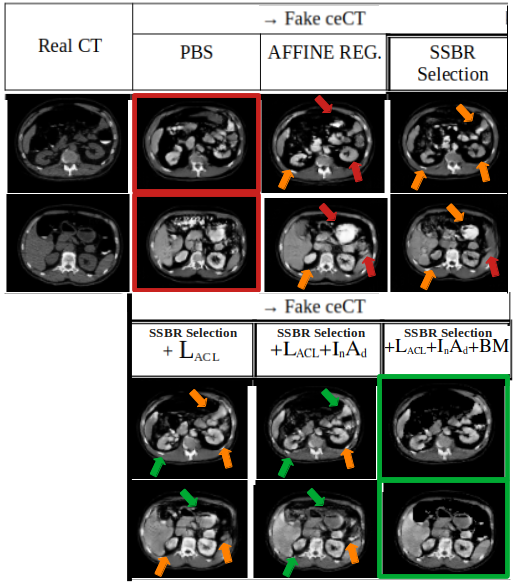}
\includegraphics[width=0.65\linewidth,height=0.43\textheight]{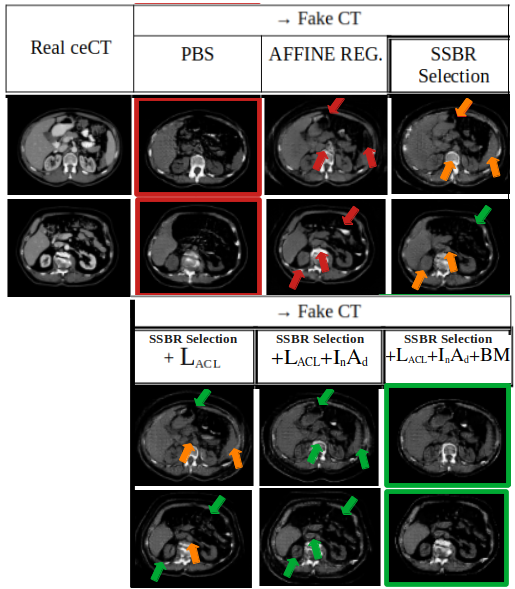}
\caption{Qualitative results on unpaired slices. Top table: CT to ceCT. Bottom table: ceCT to CT. Tests based on CycleGAN. Last four columns: results with our methods, we add each proposition to the SSBR selection. $I_nA_d$ indicates the input addition while $BM$ the use of binary mask. Arrows: high (red), low (orange) and no (green) artefacts.}
\label{fig:qualitative}
\vspace{-4mm}
\end{figure}

\noindent Despite the good results shown in the method identified as the best ones in 
Figure~\ref{fig:sota}, in terms of overall shape and contrast intensity, the PBS selection was not sufficient and several anatomical artefacts appeared (see Figure~\ref{fig:qualitative}). Another existing strategy for anatomically-paired selection that we tested was the use of 3D {\em affine} registration with Simple-Elastix~\cite{simple-eslatix} algorithm. Given the high variability between the two domains, we decided to perform the registration at each iteration between the two selected patients.
Anatomical coherence was improved 
but some important artefacts still appeared. Finally, our proposed selection with SSBR was tested, which reduced the severity of artefacts, as shown in the forth column of Figure~\ref{fig:qualitative}. 

We then \GLB{added the $L_{ACL}$ loss function, which significantly improved anatomical coherence, particularly in the binary mask regions. Eventually,} we combined in a first moment the use of input as template $I_nA_d$ and in a second moment the binary mask $BM$. 
The complete proposed method based on SSBR selection with $L_{ACL}$, $I_nA_d$ and $BM$ produced high quality synthetic images, without visual artefacts and with realistic contrast intensity \GLB{according to physicians' evaluation}. Some qualitative results are detailed in Figure~\ref{fig:qualitative}. 

\subsection{Quantitative ablation study on paired database}
\textbf{Dataset} 
For quantitative testing we used a pathological private pediatric database of paired abdominal ceCT-CT images of 10 patients with renal cancer. It is important to note that the small number of patients in this data set prevents \GLB{achieving satisfactory performance on training} generative models.No public paired ceCT-CT datases is available, so this small dataset we gathered is quite rare. Moreover, we want to emphasize that for each subject we have at our disposal about 100 2D slices of the renal ROI and therefore the results refer to a total of about 1000 2D images.\\
\textbf{Results}
The quantitative ablation and comparative study was performed using the presented methods, pre-trained on the unpaired data-sets. The results are presented in Table~\ref{table:quantitative} using mean square error (MSE), structure similarity (SSIM) and peak signal to noise ratio (PSNR) between {\em real} and {\em fake} images, and training time (TIME). All our contributions improve on the original CycleGAN with PBS~\cite{pbs}, at the cost of some additional learning time (note that the inference time remains the same for all methods). \GLB{Their combination produces the best results for both tasks.}
The use of {\em affine} registration seems to be quantitatively comparable to the use of SSBR for the selection and loss function $L_{ACL}$, but the network requires a very high computational time in addition to the creation of artefacts, as illustrated in Figure~\ref{fig:qualitative}.

\begin{table}[htbp]
 \centering
 \caption{Quantitative study on the 10 patients with paired images. Mean square error (MSE), structure similarity (SSIM) and peak signal to noise ratio (PSNR) are shown as mean and standard deviation. TIME is the training time. Tests based on CycleGAN. The first section \GLB{of rows} presents the ablation study, in which each proposition is added to the SSBR selection. $I_nA_d$ indicates the input addition while $BM$ the use of binary mask.}
 \label{tab:quantitativecyclegan}
 \resizebox{0.9\columnwidth}{!}{%
 \begin{tabular}{|c||c|c|c|c|}
 \hline
 CycleGAN Method & MSE [$10^{-2}$] ($\downarrow$) & SSIM [$10^{-1}$] ($\uparrow$) & PSNR ($\uparrow$) & TIME ($\downarrow$) 
 \\
 \hline
 \multicolumn{5}{|c|}{\centering real CT$\rightarrow$fake ceCT vs real ceCT}
 \\
 \hline
 $+ L_{ACL} + I_nA_d + BM$ & \textbf{6.37 (2.01)} & \textbf{6.81 (0.62)} & \textbf{18.14 (1.23)} & 7h 55m
 \\
 $+ L_{ACL} + I_nA_d$ & 6.41 (1.97) & 6.67 (0.63) & 18.11 (1.22) & 7h 55m
 \\
 $+ L_{ACL} + BM$ & 8.19 (2.32) & 6.36 (0.72) & 17.02 (1.14) & 7h 49m
 \\
 $+ I_nA_d$ & 6.79 (2.85) & 6.60 (0.74) & 17.97 (1.54) & 7h 14m
 \\
 $+ BM$ & 8.42 (2.46) & 6.24 (0.73) & 16.91 (1.17) & 7h 5m
 \\
 $+ L_{ACL}$ & 8.55 (2.28) & 6.19 (0.69) & 16.82 (1.07) & 7h 49m
 \\
 SSBR selection & 9.07 (2.39) & 5.99 (0.71) & 16.56 (1.07) & 7h 5m
 \\
 \hline
 AFFINE REG. & 8.16 (1.80) & 6.36 (0.57) & 16.99 (0.87) & 16h 33m
 \\
 \hline
 PBS & 10.05 (2.89) & 5.76 (0.65) & 16.14 (1.15) & 3h 2m
 \\
 \hline
 \multicolumn{5}{|c|}{\centering real ceCT$\rightarrow$fake CT vs real CT}
 \\
 \hline
 $+ L_{ACL} + I_nA_d + BM$ & \textbf{4.05 (0.83)} & \textbf{7.23 (0.53)} & \textbf{20.03 (0.92)} & 7h 55m
 \\
 $+ L_{ACL} + I_nA_d$ & 4.24 (0.86) & 6.80 (0.37) & 19.83 (0.92) & 7h 55m
 \\
 $+ L_{ACL} + BM$ & 5.08 (0.85) & 6.87 (0.52) & 19.02 (0.74) & 7h 49m
 \\
 $+ I_nA_d$ & 6.16 (1.15) & 5.87 (0.23) & 18.18 (0.79) & 7h 14m
 \\
 $+ BM$ & 6.07 (1.28) & 6.61 (0.65) & 18.28 (0.99) & 7h 5m
 \\
 $+ L_{ACL}$ & 5.87 (1.73) & 6.08 (0.22) & 18.47 (1.12) & 7h 49m
 \\
 SSBR selection & 7.15 (2.16) & 5.68 (0.52) & 17.64 (1.26) & 7h 5m
 \\
 \hline
 AFFINE REG. & 4.72 (0.95) & 6.77 (0.37) & 19.36 (0.93) & 16h 33m
 \\
 \hline
 PBS & 8.26 (1.97) & 5.36 (0.28) & 16.96 (1.04) & 3h 2m
 \\
 \hline
 \end{tabular}
 }
\vspace{-4mm}%
\label{table:quantitative}
\end{table}

\subsection{Blood vessel segmentation using ceCT and CT}
\textbf{Dataset}
For the proposed segmentation application, the synthetic images used were produced using generative methods trained as explained previously but with images at the original size 512$\times$512. Reference segmentations of arteries and veins were manually performed by medical experts on our paired pathological dataset.\\
\textbf{Segmentation performances}
To further demonstrate the realness of the images generated by our method, 
similarly to~\cite{dda-gan,sandfort2019,song2020,zhu2020}, we \PG{compared the performance of a segmentation network when using either a real image and a fake image, or both real images}.
Given the restricted dataset, all tests were done with the Leave-One-Patient-Out (L-O-P-O) method using the 3D nnU-Net~\cite{nnU-Net}. 
Results show that replacing a real CT modality with a synthetic one produced with CycleGAN and the PBS method, as in~\cite{sandfort2019,song2020}, is not sufficient to achieve performances as good as when using both real modalities. By contrast, the synthetic CT images produced by our method achieve 
the highest Dice score and the lowest Hausdorff distance, with the best combination of precision and recall. This is even more evident for the more heterogeneous cases, particularly for the veins. 
Quantitative results are shown in Table~\ref{table:segmentation}. Some qualitative results and other quantitative results for an extended ceCT dataset (no CT images available) can be found in the supplementary material.

\begin{table}[htbp]
 \centering
 \caption{Segmentation performance on \textbf{real ceCT} of 10 patients (and then on the only 5 more heterogeneous cases) using L-O-P-O methods. Dice score (DS), precision (PR), recall (RC) and 95th percentile of the Hausdorff distance (HD95) are given (mean and standard deviation). All tests were done using 3D nnU-Net~\cite{nnU-Net} with intensity (except if indicated) and geometric data augmentation.} 
 \label{tab:quantitativecyclegan}
 \resizebox{1.0\columnwidth}{!}{%
 \begin{tabular}{|c|c||c|c|c|c|}
 \hline
 INPUT Database & Structure & DS [100\%] ($\uparrow$) & PR [100\%] ($\uparrow$) & RC [100\%] ($\uparrow$) & HD95 [mm] ($\downarrow$) 
 \\
 \hline
 \multicolumn{6}{|c|}{\centering \textbf{on 10 patients}}
 \\
 \hline
 \multirow{2}{*}{real ceCT and real CT} & Arteries & 74.61 (5.89) & 85.22 (8.32) & 69.06 (8.15) & 15.39 (5.72)
 \\
 & Veins & 45.62 (13.72) & 60.61 (19.53) & 38.68 (14.83) & 31.47 (16.53)
 \\
 \hline
 \multicolumn{6}{|c|}{}
 \\
 \hline
 \multirow{2}{*}{real ceCT without data aug.} & Arteries & 63.75 (11.18) & 80.33 (10.99) & 53.88 (12.48) & 23.43 (8.18)
 \\
 & Veins & 21.18 (19.70) & 64.04 (34.08) & 15.45 (16.04) & 42.14 (23.79)
 \\
 \hline
 \multirow{2}{*}{real ceCT} & Arteries & 73.01 (6.57) & 81.08 (8.70) & 67.19 (8.43) & 15.80 (7.01)
 \\
 & Veins & 40.58 (23.50) & 55.94 (31.39) & 33.72 (26.61) & 40.65 (30.90)
 \\
 \hline
 \multirow{2}{*}{real ceCT and fake$_{PBS}$ CT} & Arteries & 69.59 (8.89) & 79.54 (10.85) & 63.47 (12.59) & 18.08 (8.21)
 \\
 & Veins & 44.40 (22.75) & 58.44 (21.78) & 38.38 (23.20) & 39.31 (16.79)
 \\
 \hline
 \multirow{2}{*}{real ceCT and fake$_{Ours}$ CT} & Arteries & \textbf{72.33 (7.41)} & 77.29 (10.32) & 68.63 (8.88) & \textbf{15.48 (6.38)}
 \\
 & Veins & \textbf{44.49 (22.50)} & 54.98 (26.74) & 40.28 (22.69) & \textbf{38.90 (32.76)}
 \\
 \hline
 \multicolumn{6}{|c|}{\centering \textbf{on 5 more heterogeneous}}
 \\
 \hline
 \multirow{2}{*}{real ceCT and real CT} & Arteries & 75.01 (5.82) & 85.17 (4.37) & 67.50 (8.57) & 12.79 (6.04)
 \\
 & Veins & 40.87 (14.73) & 56.93 (18.63) & 32.62 (13.05) & 31.16 (10.76)
 \\
 \hline
 \multicolumn{6}{|c|}{}
 \\
 \hline
 \multirow{2}{*}{real ceCT without data aug.} & Arteries & 66.59 (8.31) & 86.89 (5.70) & 54.83 (10.29) & 23.34 (9.14)
 \\
 & Veins & 14.66 (17.05) & 71.31 (39.90) & 8.89 (10.98) & 50.35 (29.50)
 \\
 \hline
 \multirow{2}{*}{real ceCT} & Arteries & 72.94 (6.30) & 84.37 (3.80) & 64.89 (9.71) & 13.49 (5.14)
 \\
 & Veins & 28.28 (19.84) & 51.97 (38.06) & 17.50 (18.41) & 35.57 (14.33)
 \\
 \hline
 \multirow{2}{*}{real ceCT and fake$_{PBS}$ CT} & Arteries & 70.77 (9.18) & 84.41 (5.96) & 63.00 (15.51) & 13.83 (5.95)
 \\
 & Veins & 33.47 (26.92) & 45.48 (34.33) & 27.73 (23.78) & 37.73 (23.42)
 \\
 \hline
 \multirow{2}{*}{real ceCT and fake$_{Ours}$ CT} & Arteries & \textbf{73.18 (7.51)} & 80.58 (4.59) & 67.63 (11.25) & \textbf{12.73 (4.10)}
 \\
 & Veins & \textbf{40.57 (20.25)} & 62.01 (13.31) & 31.96 (18.91) & \textbf{32.83 (13.84)}
 \\
 \hline
 \end{tabular}
 }
\vspace{-4mm}%
\label{table:segmentation}
\end{table}

\section{Conclusion}
\label{sec:conc}
We presented an extension of CycleGAN via the use of a Self-Supervised Body Regressor to: (i) better select anatomically-paired slices; (ii) anatomically constrain the generator to produce a slice describing the same anatomical content as the input. We applied our method to the unsupervised synthesis of ceCT-CT images. We \HBO{showed} significant improvements in the generated images compared to existing methods. To further validate our method, we demonstrated that the synthesized images can be used to guide a segmentation method by compensating, without loss of performance, for the absence of the complementary real acquisition modality.
\GLB{Future work aims to apply our method on other translation tasks, such as MRI to CT or T1-w to T2-w, and other body sections. Moreover, once more data and more powerful GPUs will be available, $transformer$-based methods will be futher explored.}

\paragraph*{Acknowledgments}
This work has been partially funded by a grant from Region Ile de France (DIM RFSI).

\newpage
\appendix
\section{Supplementary material}

\begin{figure}[htbp]
\begin{minipage}[b]{1.0\linewidth}
  \centering
\includegraphics[width=0.55\linewidth]{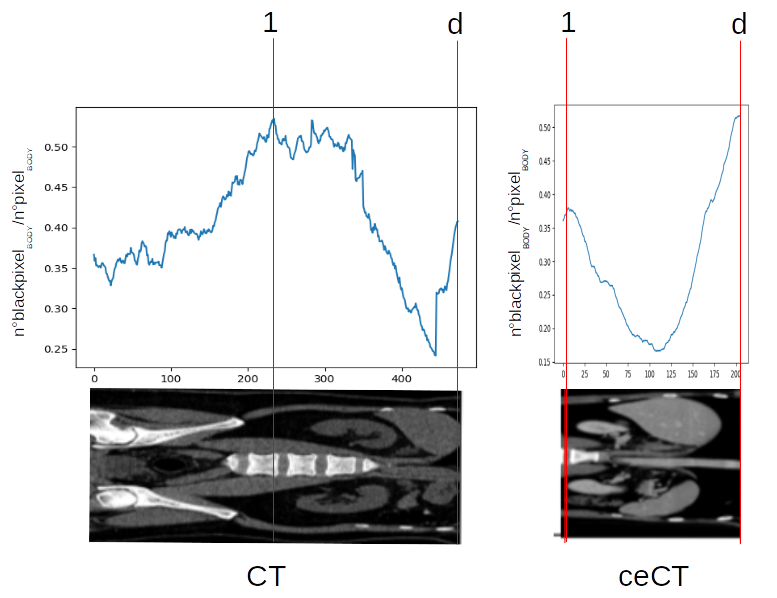}
\includegraphics[width=0.8\linewidth]{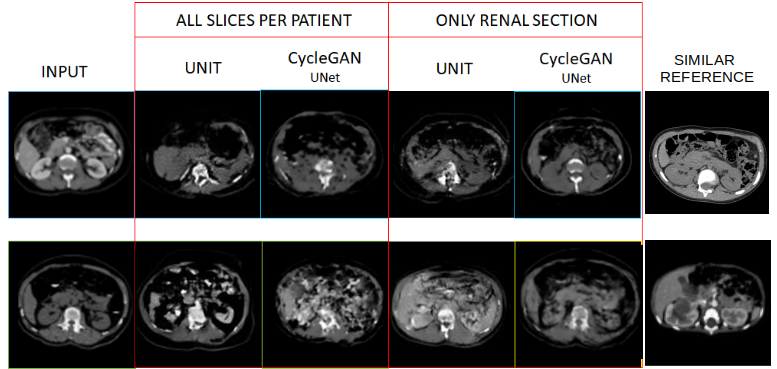}

\caption{Top: example of renal ROI selection using the number of black pixel in relation to the total pixels of each individual slice to automatically select the first slice of the lungs and the last slice of the visceral area as the upper and lower reference points respectively. Bottom: comparison of CycleGAN and UNIT trained without and with renal ROI selection (no PBS strategy used). For both methods, we used U-Net as generator network and Patch-GAN as discriminator mechanism. First row: from ceCT to CT. Second row: from CT to ceCT. An idea of how the expected output should look like is provided in the last column.}
\label{fig:no-landmark}
\end{minipage}
\end{figure}

\begin{figure}[htbp]
  \centering
\includegraphics[width=1.0\linewidth]{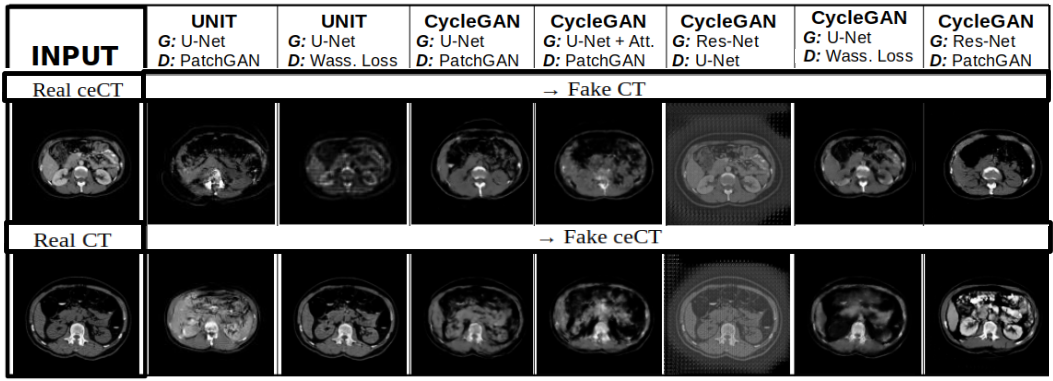}
\caption{Comparison of some other state-of-the-art methods on slices of the unpaired test set. 1st row: ceCT to CT. 2nd row: CT to ceCT. The slices in all tests are selected with PBS in the renal section. The input in the other direction gives an idea of what the expected result should look like. ``Att." indicates the use of an attention layer as the last layer, while ``Wass. Loss" the use of Wasserstein distance as Discriminator Loss.}
\label{fig:sota}
\end{figure}

\begin{table}[htbp]
  \centering
  \caption{Some of the parameters for 3D $affine$ registration using SimpleITK-SimpleElastix} 
  \resizebox{\columnwidth}{!}{%
\begin{tabular}{|c|c|}
    \hline
    \textbf{Parameter Name} & \textbf{Parameter Value}  \\
    \hline
    \hline
  Final BSpline Interpolation Order & 2 \\
      \hline
  Interpolator & Linear Interpolator \\
      \hline
  Maximum Number Of Iterations & 32 \\
      \hline
  Maximum Number Of Sampling Attempts & 8 \\
      \hline
  Metric & Advanced Mattes Mutual Information \\
      \hline
  Number Of Samples For Exact Gradient & 4096 \\
      \hline
  Number Of Spatial Samples & 4096 \\
      \hline
  Optimizer & Adaptive Stochastic Gradient Descent \\
      \hline
  Registration & Multi Resolution Registration \\
      \hline
  Resample Interpolator & Final BSpline Interpolator \\
    \hline
  \end{tabular}
  }
\label{table:simple-elastix}
\end{table}

\begin{figure}[htbp]
  \centering
\includegraphics[width=1.0\linewidth]{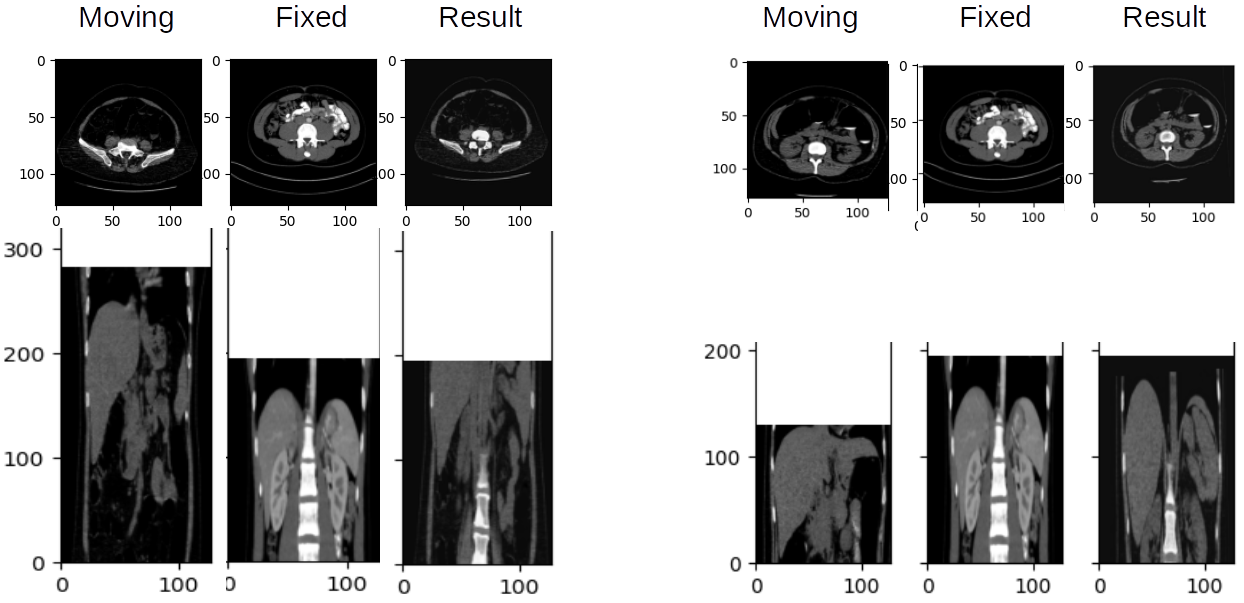}
\caption{Two examples of 3D $affine$ registration using SimpleElastix. }
\label{fig:affine_example}
\end{figure}

\begin{figure}[htbp]
  \centering
\includegraphics[width=1.0\linewidth]{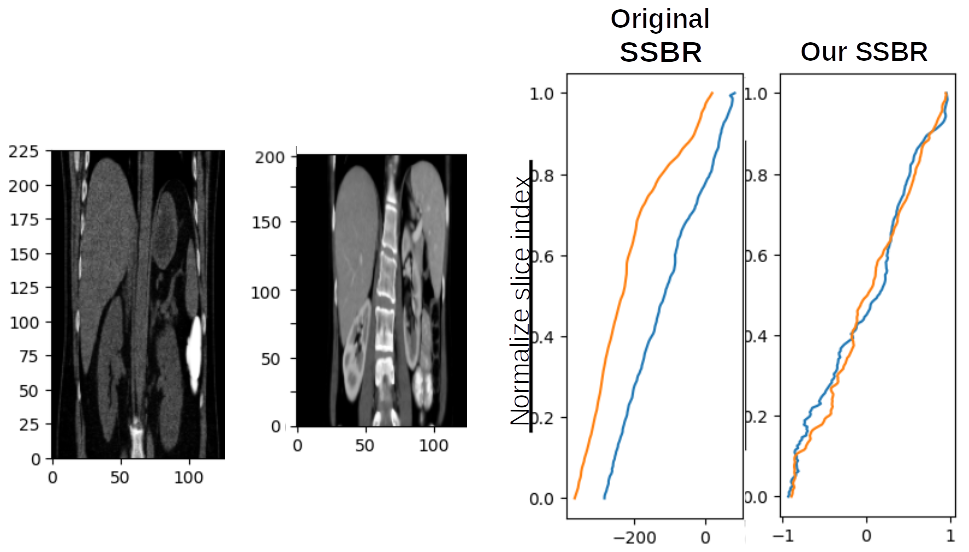}
\includegraphics[width=0.8\linewidth]{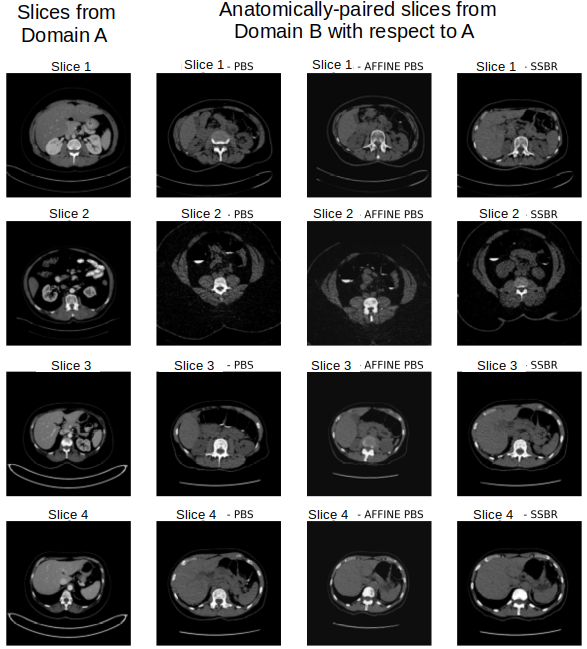}
\caption{Top: example to show differences in SSBR scores using the original training and our proposed one. Bottom: some examples of input selection methods. Image B is chosen starting from image A with a Position-Based Selection ($PBS$), 3D $affine$ registration+$PBS$ or our proposed SSBR selection.}
\label{fig:ssbr}
\end{figure}

\begin{figure}[htbp]
  \centering
\includegraphics[width=0.85\linewidth]{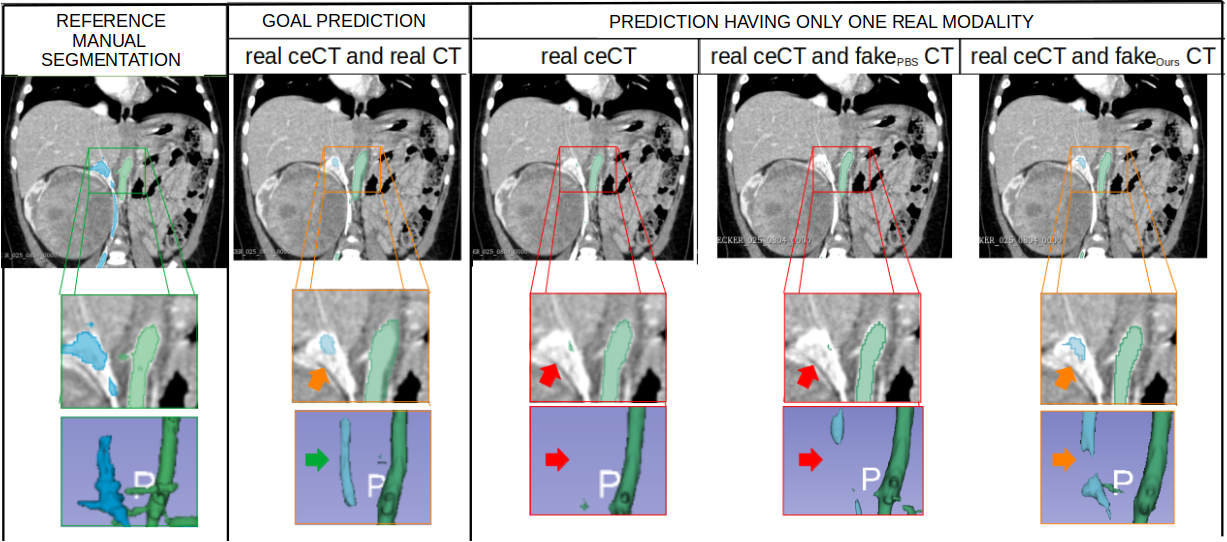}
\includegraphics[width=0.85\linewidth]{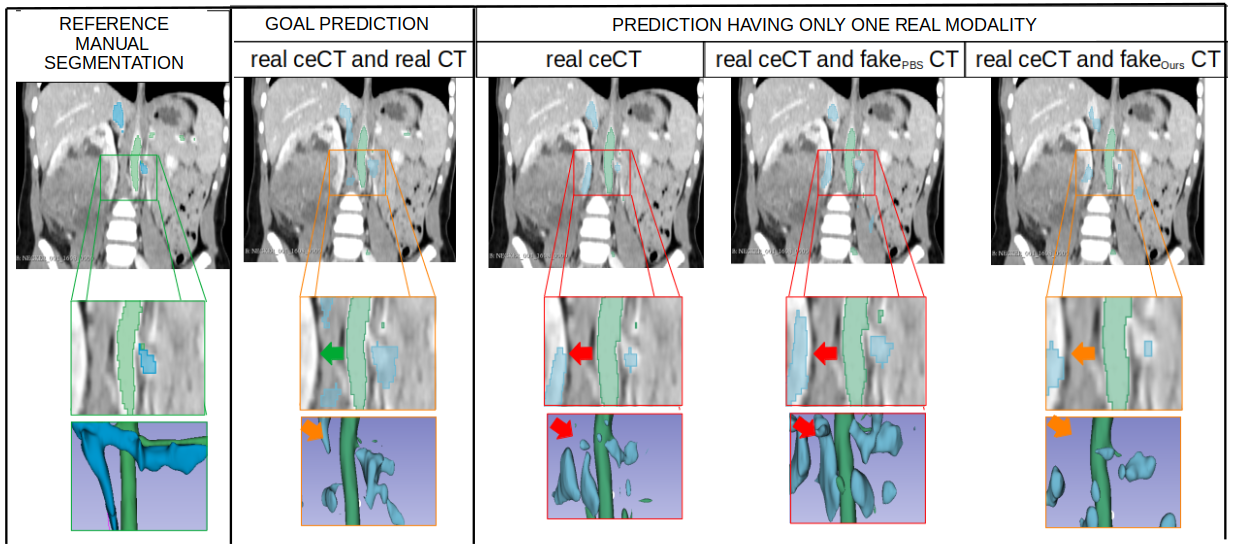}
\caption{Segmentation results of the most heterogeneous patient (top) 
and the least heterogeneous one (bottom). 
Arteries are displayed in green, and veins in blue. Arrows: strong (red), light (orange) and no (green) error.}
\label{fig:qualitativeceCT}
\end{figure}

\begin{table}[htbp]
  \centering
  \caption{Segmentation performance on \textbf{real ceCT} of 15 patients of an \textbf{extended ceCT private dataset}, composed of 65 patients (we used 43 for training and 7 for validation). Dice score (DS), precision (PR), recall (RC) and 95th percentile of the Hausdorff distance (HD95) are given (mean and standard deviation). All tests were done using 3D nnU-Net framework with intensity and geometric data augmentation.} 
  \resizebox{\columnwidth}{!}{%
\begin{tabular}{|c|c||c|c|c|c|}
    \hline
    INPUT Database & Structure & DS [100\%] ($\uparrow$) & PR [100\%] ($\uparrow$) & RC [100\%]  ($\uparrow$) & HD95 [mm] ($\downarrow$)
    \\
    \hline
    \multicolumn{6}{|c|}{\centering \textbf{on 15 patients}}
    \\
    \hline
    \multirow{2}{*}{real ceCT} & Arteries & 63.45 (5.67) & 71.73 (9.99) & 57.87 (7.31) & 17.46 (9.65)
    \\
    & Veins & 42.64 (20.12) & 76.67 (13.17) & 31.84 (17.12) & 23.55 (17.00)
    \\
    \hline
    \multirow{2}{*}{real ceCT and fake$_{PBS}$ CT} & Arteries & 65.60 (4.45) & 73.04 (10.83) & 60.91 (7.12) & 15.59 (8.47)
    \\
    & Veins & 45.77 (18.67) & 73.14 (14.88) & 35.37 (17.87) & 21.25 (20.05)
    \\
    \hline
    \multirow{2}{*}{real ceCT and fake$_{Ours}$ CT} & Arteries & \textbf{70.01 (3.99)} & 76.29 (8.23) & 65.77 (7.73) & \textbf{13.47 (10.09)}
    \\
    & Veins & \textbf{56.55 (20.20)} & 81.53 (8.91) & 46.98 (22.38) & \textbf{20.93 (22.96)}
    \\
    \hline
    \multicolumn{6}{|c|}{\centering \textbf{on 5 more heterogeneous}}
    \\
    \hline
    \multirow{2}{*}{real ceCT} & Arteries & 63.23 (4.24) & 74.86 (7.53) & 54.99 (4.55) & 15.54 (6.08)
    \\
    & Veins & 27.43 (20.62) & 66.64 (15.59) & 19.90 (17.58) & 24.90 (8.42)
    \\
    \hline
    \multirow{2}{*}{real ceCT and fake$_{PBS}$ CT} & Arteries & 64.97 (1.12) & 76.68 (11.92) & 57.61 (5.97) & 15.49 (5.26)
    \\
    & Veins & 33.16 (18.83) & 62.77 (18.28) & 24.18 (16.09) & 20.91 (7.55)
    \\
    \hline
    \multirow{2}{*}{real ceCT and fake$_{Ours}$ CT} & Arteries & \textbf{70.15 (3.52)} & 80.40 (9.97) & 62.89 (4.71) & \textbf{12.15 (6.65)}
    \\
     & Veins & \textbf{37.00 (16.11)} & 77.58 (11.78) & 26.01 (14.02) & \textbf{21.71 (6.33)}
    \\
    \hline
  \end{tabular}
  }
\label{table:segmentation4}
\end{table}

\newpage
\bibliography{refs}

\end{document}